# Analysis of a Casimir-driven Parametric Amplifier with Resilience to Casimir Pull-in for MEMS Single-Point Magnetic Gradiometry

Josh Javor[1]*, Zhancheng Yao[2], Matthias Imboden[3], David K. Campbell[2,4,5] and David J. Bishop[1,2,4,5,6]

1. Department of Mechanical Engineering, Boston University, Boston, MA 02215
2. Division of Materials Science and Engineering, Boston University, Boston MA 02215
3. 4K-MEMS, St. Blaise, Switzerland
4. Department of Electrical Engineering, Boston University, Boston, MA 02215
5. Department of Physics, Boston University, Boston, MA 02215
6. Department of Biomedical Engineering, Boston University, Boston, MA 02215

**Abstract:**

The Casimir Force, a quantum mechanical effect, has been observed in several microelectromechanical systems (MEMS) platforms. Due to its extreme sensitivity to the separation of two objects, the Casimir Force has been proposed as an excellent avenue for quantum metrology. Practical application, however, is challenging due to attractive forces leading to stiction and failure of the device, called Casimir pull-in. In this work, we design and simulate a Casimir-driven metrology platform, where a time-delay based parametric amplification technique is developed to achieve a steady state and avoid pull-in. We apply the design to the detection of weak, low frequency, gradient magnetic fields, similar to those emanating from ionic currents in the heart and brain. Simulation parameters are selected from recent experimental platforms developed for Casimir metrology and magnetic gradiometry, both on MEMS platforms. While MEMS offer many advantages to such an application, the detected signal must typically be at the resonant frequency of the device, with diminished sensitivity in the low frequency regime of biomagnetic fields. Using a Casimir-drive parametric amplifier, we report a 10,000 fold improvement in the best-case resolution of MEMS single-point gradiometers, with a maximum sensitivity of 6 Hz/(pT/cm) at 1 Hz. The development of the proposed design has the potential to revolutionize metrology, and specifically may enable unshielded monitoring of biomagnetic fields in ambient conditions.

**Introduction:**

Quantum fluctuations in the electromagnetic field give rise to forces between conductors at the same potential, when their separation is near 100 nm. Known as the Casimir force, this phenomenon was first predicted by H. B. G. Casimir[1], and was later expanded to arbitrary materials[2,3]. Since then, the Casimir Force has been measured experimentally many times[4-11], and has been proposed as a practical metrology platform using micro- and nano- electromechanical systems (MEMS/NEMS)[4-6,12]. The Casimir Force is attractive for metrology applications due to its extreme sensitivity to the separation between two objects and the ability to be measured in ambient conditions[5]. One of the most critical challenges in the development of practical platforms is the resilience to Casimir pull-in, which results in stiction in MEMS/NEMS devices and typically irreversible damage[13,14]. Notably, Casimir pull-in is only an obstacle in devices which employ an attractive force[4-6,8-11,13], but a specific combination of materials can generate a repulsive force as well[7,15]. The repulsive force configuration, however, has only been observed in liquid, which is not ideal for many common MEMS/NEMS applications, such as resonant sensing. As such, many have proposed avoiding pull-in in attractive force platforms by optimizing the material dielectric properties[16], the radius of curvature in a sphere-plate configuration[17], and the roughness of interacting surfaces[18,19]. A device capable of leveraging the attractive Casimir Force while resilient to pull-in would realize great utility for quantum metrology.

Quantum metrology has long-standing application to the measurement of very weak magnetic fields. The superconducting quantum interference device (SQUID) measures changes in magnetic field associated with a flux quantum[20]. The atomic magnetometer (AM) measures a quantum effect involving the magnetic spin states of atoms in a vapor cell[21]. The list of applications for such high resolution magnetic sensors is vast, spanning biomagnetic detection of cardiac contractions[22-23], electromagnetic brainwaves[24], and solid cancerous tumors[25], to astronomical observations, such as Jupiter's magnetosphere[26]. Sensitive magnetometry is often challenged by interference from ambient geomagnetic field and nearby electromagnetic sources. For real-time measurements, the most common methods to reduce effects of interference are magnetic shielding and gradiometry. Shielding is both expensive and cumbersome, and so there is a great effort to conduct unshielded measurements, where gradiometry and sensor design are critical[23-24]. Gradiometry involves the measurement of the gradient of the magnetic field, as opposed to the field itself, to reduce interference from nearby sources. A gradiometric measurement is typically achieved by the subtraction of two magnetometers, and was recently achieved in a single-point measurement[27] at a resolution in the range of magnetocardiography (MCG). The application of gradiometric methods have been shown to reduce the need for shielded environments in sensitive measurements, such as MCG[23-24].

Existing gradiometer technology is reviewed in **Fig. 1**, repurposed from ref. 27 for this work. The fields are illustrated in gradients, assuming the sensors of each technology can be separated by 1 cm. All technologies involve subtraction between two sensors to calculate the gradient, except the MEMS single-point gradiometer[27]. The first order gradients of the geomagnetic field (E, 500 pT/cm) and geomagnetic noise (GMN, 500 fT/cm) are illustrated by dash-dotted lines. SQUID and AM technology clearly lead the group in resolution. Myriad other techniques are used to sense magnetic fields, where the bottom band (in purple) are MEMS designs. MEMS magnetic sensors are typically resonant sensors with high quality factors, where impressive sensitivity has been achieved at higher frequencies in the kHz range[27-29]. In order to detect low frequency biomagnetic fields, nontrivial techniques would be necessary to employ resonant sensors.

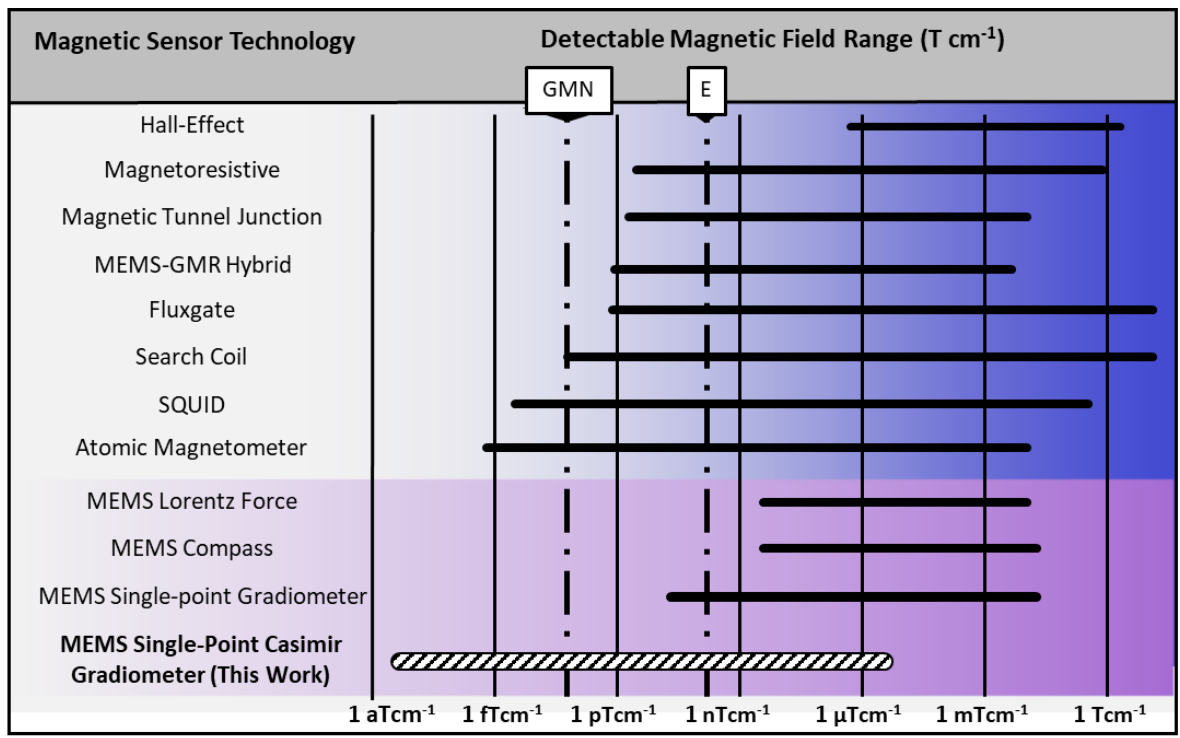

**Fig. 1: Magnetic Gradiometer Technology.** All magnetic fields are illustrated as gradients. Existing sensor technology is adjusted by taking the best resolution in literature and assuming two sensors can be separated by 1 cm, which can measure a gradient field. E is Earth's field gradient and GMN is the gradient geomagnetic noise. The devices shaded in purple at the bottom are microelectromechanical systems (MEMS) technology. Our previous MEMS single-point gradiometer design[27] experimentally achieved a resolution of 100 pT/cm, with a calculated best-case resolution of 30 fT/cm at resonance. This work (black, dashed) proposes a quantum-derived enhancement to this design, which suggests a 10,000 fold improvement in the best-case resolution with a measurement near DC. This figure is repurposed from ref. 27.

The sensitivity of resonant MEMS devices can be tuned by many techniques, such as parametric amplification[12,30-31] and modal coupling[32-33]. Conceptually, parametric amplification is typically achieved by modulating a parameter of the equation of motion at two times the frequency of resonance and control of the phase relation between the driven mode and the modulated parameter. When applied to the drive system of a MEMS cantilever[30], thermomechanical noise was reduced in one phase by 4.9 dB. A Lorentz force MEMS magnetometer was parametrically driven to enhance sensitivity at resonance by over 80 fold[31]. The gain in such electrostatic systems is typically 10-1000. Parametric modulation can also be applied to coupled resonators, where one object, or mode, oscillates at twice the frequency of another. Parametric pumping was shown to dynamically tune the coupling of two modes in a gyroscopic ring resonator, with application in inertial sensing[32]. In the design of a Casimir-coupled resonator[12], a gold sphere was used to parametrically pump the oscillation of a torsional oscillator, proposed to amplify a DC voltage measurement up to ten orders of magnitude. Using an attractive Casimir Force design, the challenge of avoiding Casimir pull-in was also discussed in depth. Tunability of such systems is often critical for experimental utility, and to account for small fabrication asymmetries, which have a significant impact on the coupling[12,32].

In this work, we propose the coupling between two resonators by the quantum-derived Casimir Force. Fundamentally, we combine two metrology platforms, a MEMS Casimir platform, and a MEMS single-point gradiometer platform. The Casimir platform[5] comprises a MEMS accelerometer functionalized by a gold sphere, where the voltage potential can be controlled. A Casimir Force measurement is achieved when a gold plate is brought within 100 nm of the sphere. The MEMS gradiometer platform[27] comprises a MEMS accelerometer functionalized by a cube micromagnet. Measurement is achieved by detecting an oscillating force on the permanent magnet, at resonance, where the force is proportional to a gradient magnetic field. Both resonators can also be driven in analog electrostatically[5,27,34]. In essence, these two systems on the same central axis and within 100 nm separation form a Casimir-coupled resonator. Similar to the Casimir oscillator using a torsional plate[12], this design would be highly sensitive to small changes in the DC separation, due to the Casimir coupling. Therefore, DC forces from gradient magnetic fields will be intensely amplified by the coupling. Then, instead of using parametric amplification to further amplify the sensitivity, we design a technique to achieve a steady state Casimir oscillator, resilient to pull-in. In this work, we propose and simulate a highly tunable Casimir-driven Gradiometer, sensitive to slowly varying magnetic fields and resilient to pull-in.

**MEMS Casimir Gradiometer Design**

The design in this work is a nontrivial combination of an experimental MEMS gradiometer platform[27], an experimental MEMS Casimir force metrology platform[5], and a coupling method[12] to parametrically modulate the interaction between the two platforms. In order to approach the divergent gain of this physical design, we anticipate that a tunable experimental platform with precise positional control of the two microobjects (magnet and sphere) is required. When the separation is reduced to near 100 nm, the Casimir force contribution becomes significant, as has been shown previously[4,5]. If the separation decreases much past this, the system is predicted to experience Casimir pull-in, an event that is caused by the attractive Casimir force overcoming the restoring force of the spring, and causing the device to malfunction.

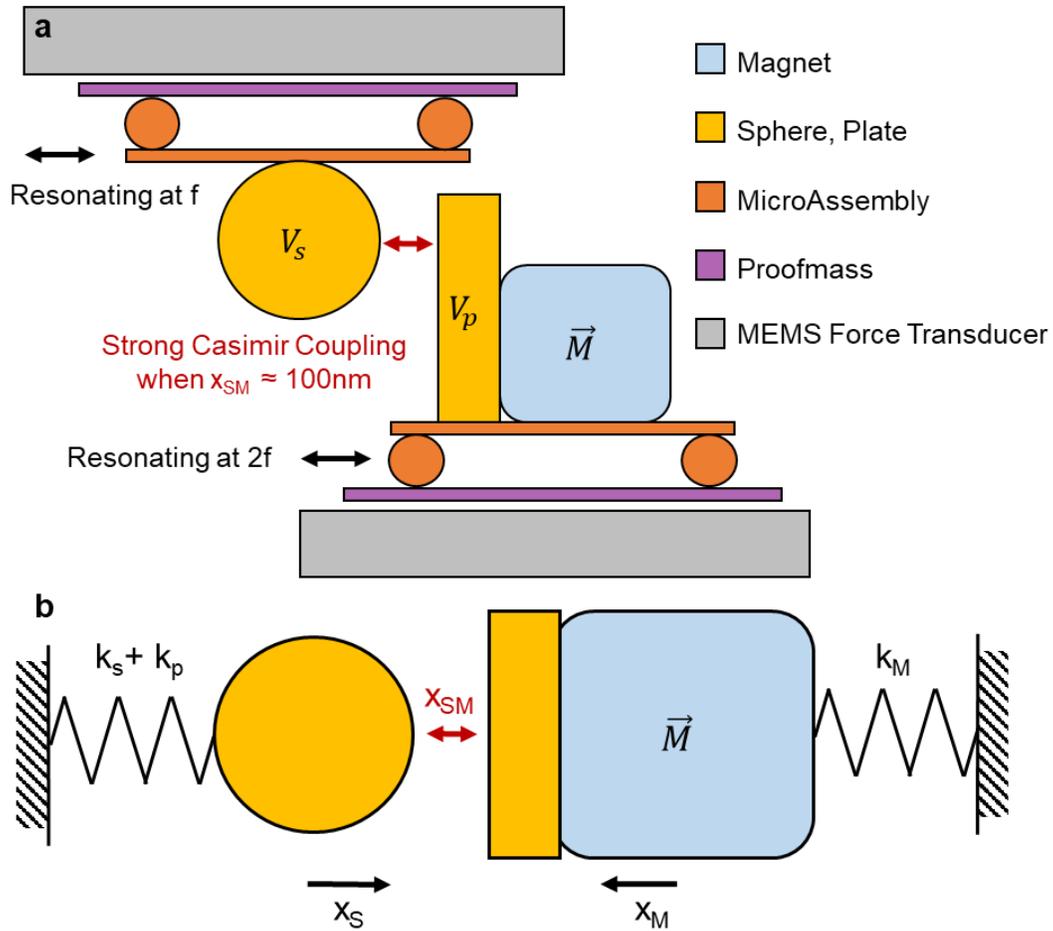

**Fig. 2: Proposed MEMS Casimir Gradiometer Design.** The free body diagram (a) shows a sphere positioned in close proximity to a magnet, where the resonant modes are coupled by the quantum-derived Casimir Force. Both objects are driven at resonance, where the micromagnet is driven at two times the frequency of the microsphere, to accomplish parametric modulation. Force from a weak gradient magnetic field will induce a small deflection of the magnet, decreasing the separation, $x_{SM}$, and inducing a frequency and amplitude shift in the sphere oscillation, via the Casimir coupling. The proposed experimental design (b) and the accompanying parameters are chosen based on existing experimental platforms[5,27]. In order to successfully couple these two systems, a highly tunable design is necessary. By assembling the sphere and magnet on independent MEMS force transducer platforms, their independent actuation is completely controlled. By inverting the sphere platform, a separation between the sphere and plate of near 100 nm can be achieved.

The proposed sensing platform is illustrated in **Fig. 2a**. Each device is fabricated individually as described previously[5,27] using commercially available MEMS accelerometers. This is accomplished using precise control of a vacuum pick-and-place system, and feedback from a live sensor (post-release MEMS). The micromagnet used in the gradiometer design will be functionalized with a gold plate on its side that faces the microsphere. The sensor with the microsphere will be inverted such that the sphere will have a clear pathway to come into close proximity to the plate. Both platforms enable control of static position, oscillation amplitude, frequency, phase, and detection. The drive parameters are controlled via a built-in electrostatic self-test feature on the ADXL 203 platform, which can be used for analog control of the microobjects via pulsed width modulation (PWM)[34]. This feature will also be used to calibrate for and negate the effect of an anticipated electrostatic coupling between the plate and sphere, as has been shown previously[5].

The functional diagram of this design, shown in **Fig. 2b**, is based on the interaction of two resonators. The first resonator includes an attached gold microsphere and the second includes the micromagnet/gold mirror assembly. Independently, both resonators act as damped, driven harmonic oscillators. The coupling of the two resonators is based on the Casimir Force, which is dependent on their separation. Assuming that out-of-plane forces are minimal, we analyze a uniaxial system of equations (Eqs 1-3) along the central axis of the magnet.

$$m_S \ddot{x}_S + \frac{m_S \omega_{0S}}{Q_S} \dot{x}_S + k_{0S} x_S = F_{Dr-S}(t) + F_{Cas}(x_{SM}, t) \quad (1)$$

$$m_M \ddot{x}_M + \frac{m_M \omega_{0M}}{Q_M} \dot{x}_M + k_{0M} x_M = F_{Dr-M}(t) + F_{Cas}(x_{SM}, t) + F_M(t) \quad (2)$$

$$x_{SM}(t) = s_0 + x_S(t) - x_M(t) \quad (3)$$

Here, Eq 1 relates to the sphere, Eq 2 to the magnet, and Eq 3 to their separation. In Eq 1, $x_S$ is the displacement of the sphere, $m_S$ is the mass, $\omega_{0S}$ is the natural frequency, $Q_S$ the quality factor, and $k_{0S}$ the unperturbed spring constant. The sphere is driven electrostatically by $F_{Dr-S}$ and experiences an attractive coupling force, $F_{Cas}$, when the separation is small (order 100 nm). Similarly, in Eq 2, $x_M$ is the displacement of the magnet, $m_M$ is the mass, $\omega_{0M}$ is the natural frequency, $Q_M$ the quality factor, and $k_{0M}$ the spring constant. The magnet is also driven electrostatically by a force, $F_{Dr-M}$, and experiences an equal and opposite coupling force to the sphere, $F_{Cas}$, at small separations. In addition, a slow, time-varying gradient magnetic field would impose a force, $F_M$. in Eq 3, $x_{SM}$ is the real-time separation distance between the sphere and magnet, and $s_0$ is the separation in the absence of the Casimir coupling force. The forcing terms are expanded in Eqs 4-7 below.

$$F_{Dr-S}(t) = k_{0S} A_S \sin(\omega_{0C}(t + \tau_{1f})) \quad (4)$$

$$F_{Dr-M}(t) = k_{0S} A_M \sin(2\omega_{0C}(t + \tau_{2f})) \quad (5)$$

$$F_M = M \left(\frac{dB}{dx}\right) \cos(\theta) \quad (6)$$

$$F_{Cas} = -\frac{\pi^3 \hbar c}{360} \frac{R}{x_{SM}^3} \quad (7)$$

| Parameter | Symbol | Nominal Value |
|---|---|---|
| Uncoupled Resonant Frequency | $w_{0S}$ | 1 kHz |
| Quality Factor (both resonators) | $Q$ | 1000 |
| Sphere Spring Constant | $k_{0S}$ | 25 mN/m |
| Magnet Spring Constant | $k_{0M}$ | 25 mN/m |
| Sphere Mass | $m_S$ | 1 µg |
| Sphere Radius | $R$ | 60 µm |
| Sphere AC Amplitude | $A_S$ | 20 nm |
| Sphere Time Delay | $\tau_{1f}$ | 1.02 ms |
| Magnet AC Amplitude | $A_M$ | 10nm |
| Magnet Time Delay | $\tau_{2f}$ | 750µs |
| Separation | $s_0$ | ~100 nm |

**Table 1: Parameter Inputs for Simulation of Casimir Gradiometer Design.** The values are largely chosen based on reasonable implementation to existing experimental platforms[5,27]. Although not directly an input to the simulation, the typical mass, $m_M$, of the cube micromagnet in this design (with 250 µm side length) is 150 µg (ref. 27).

In Eq 4, $A_S$ is forcing amplitude of the sphere, $\omega_{0C}$ is the natural frequency of the coupled system, and $\tau_{1f}$ defines the fixed starting time. Similarly, in Eq 5, $A_M$ is the forcing amplitude of the magnet, and $\tau_{2f}$ is the time delay of magnet actuation. Most works[12,30-31] involving parametric pumping use a phase delay, and we describe our rationale for a time delay system later, in conjunction with **Fig. 4**. The magnet is forced at $2\omega_{0C}$ for parametric amplification. Eq 6 describes the force on the magnet from a gradient magnetic field predominantly along its polarized axis, where $M$ is the moment, $B$ is the magnetic field intensity, and $\theta$ is the angle between them[27]. The Casimir force for a sphere-plate geometry is displayed in Eq. 7, where $\hbar$ is Plank's constant, $c$ is the speed of light, $R$ is the radius of the sphere, and $x_{SM}$ is the separation between the sphere and plate as defined above. It should be noted that Eq. 7 is ideal (assuming absolute zero temperature and perfectly smooth infinitely conducting surfaces) and should be modified to reflect real experimental conditions[4-11] such as temperature, roughness, and dielectric properties. The nominal values for simulation inputs are shown in **Table 1**, where the primary tuning parameters are indicated. The magnet's mass (typically 150 µg, ref. 27) is not a direct input to the simulation as the magnet's dynamics are controlled by feedback (discussed later in **Fig. 4**).

When the two resonators are coupled as described in Eqs 1-3, there is a spring softening effect analogous to the electrostatic spring softening observed in capacitive systems[30]. The coupling may then be modulated by tuning the parameters of the magnet resonator. This will be necessary to access the most sensitive region of parameter space, while preventing Casimir pull-in. Following the analytical model outlined earlier[12], the equation of motion for the sphere (Eq 1) in a Casimir coupled system then becomes Eq 8 below, where the parametric spring constant, $k_p$, is defined in Eqs 9 and 10. For simplicity, we maintain the assumption of a linear spring model as the amplitudes of oscillation are small (<100 nm).

$$m_S \ddot{x}_S + \frac{m_S \omega_{0C}}{Q_C} \dot{x}_S + [k_{0S} + k_p(x_{SM}, t)] x_S = F_{Dr-S}(t) + F_{Cas}(x_{SM}, t) \quad (8)$$

$$k_p(s_0, t) = \frac{dF_{Cas}(x_{SM}, t)}{dx_{SM}} \quad (9)$$

$$k_p(s_0) = \frac{\pi^3 \hbar c}{120} \frac{R}{x_{SM}^4} \quad (10)$$

The effect of the modulated spring constant may be best illustrated by the potential energies of a quasi-static system (**Fig. 3**). The magnet is fixed (**Fig. 3a**) and the sphere, connected to a spring, is moved to set the gap between it and the conducting plate. The gap is the Casimir cavity size, $s_0$, which is defined by the equilibrium position of the spring in the absence of the Casimir Force. This has been described earlier[4] for a Casimir oscillator where $s_0$ is constant. In this work, the cavity, $s$, is influenced by movement of both the magnet and spring. The overall potential energy curve is the summation of the elastic potential energy of the spring and the potential energy of the Casimir Force, as a function of the sphere displacement, $x_S$, from the equilibrium position. At greater displacements, the Casimir attraction overcomes the restoring force, and pull-in occurs as the overall potential decreases rapidly. The overall potential energy curve is shown for varying Casimir cavity sizes (**Fig. 3b**), where each curve (from red to blue) represents the cavity decreasing by 1 nm (this is quasistatic, so the cavity size is represented by $s$). Now that the quasistatic system is understood, it is straightforward that the potential energy curve is changing dynamically as the sphere and magnet both move in real time (a dynamic cavity is represented by $x_{SM}$). Thus, with appropriate tuning of the magnet oscillation, one can modulate the sphere oscillation via the Casimir coupling. Furthermore, one can make the trade-off to tune the stability and sensitivity of the system as a larger amplitude pushes the sphere close to pull-in, and it turns the system more unstable but more sensitive to small perturbations on the magnet.

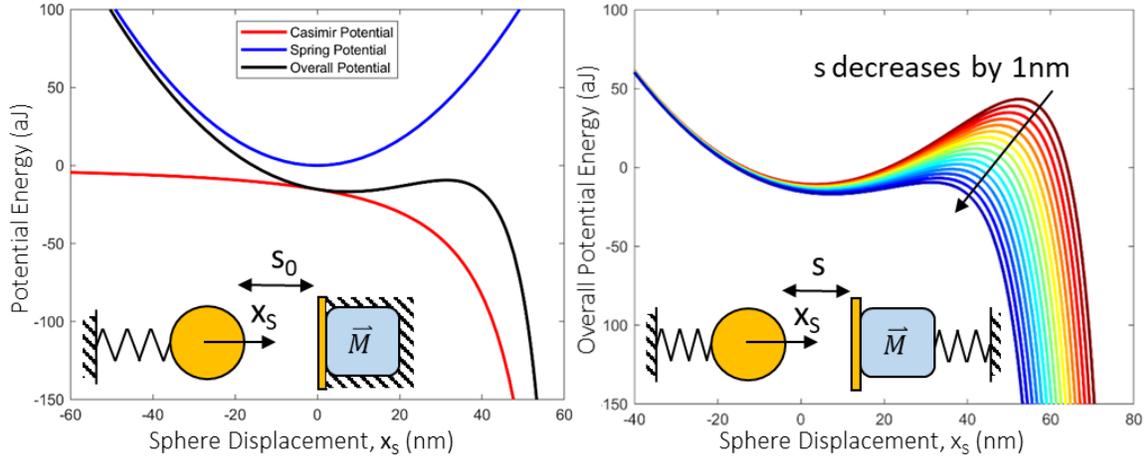

**Fig. 3: Quasistatic Illustration of Casimir Coupling** Inset: Diagram of a simple, nonlinear Casimir oscillator with (a) a fixed magnet and (b) an oscillating magnet adhered to a conducting plate. (a) Potential energy of the sphere spring without the Casimir Force (blue), and related to the Casimir Force (red). The overall potential energy (black) is shown as a function of sphere position, $x_S$. The separation $s_0$ denotes the initial separation when the spring is at equilibrium in absence of the Casimir Force. Casimir oscillators with a fixed plate have been discussed previously[4,12]. (b) Overall potential energy of sphere spring with a dynamic cavity size $s$, modulated by movement of the magnet (decreases 1 nm per curve from red to blue). In this work, the cavity is highly dynamic as both the sphere and magnet oscillate, and so we use $x_{SM}$ in the text to represent the real-time distance between the sphere and plate.

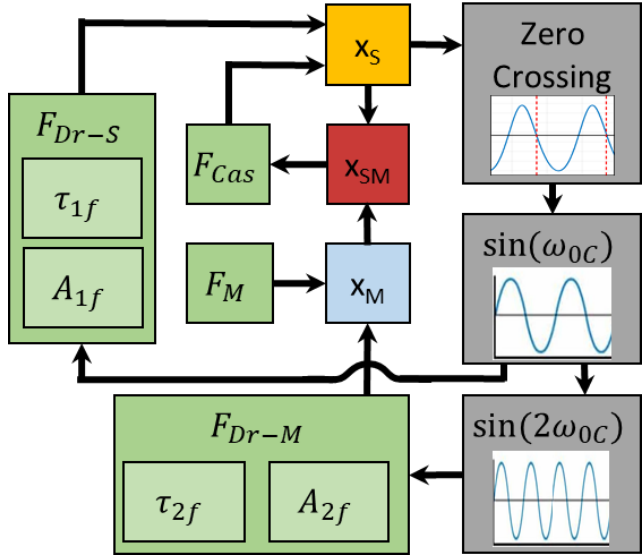

**Fig. 4: Simulation Schematic.** The simulation is designed based on instruments readily available in a standard electronics laboratory. Following Eqs 1-7 in the text, a feedback scheme is employed to tunably synchronize the f and 2f drive waveforms. The detailed schematic can be found in **Fig. S1**.

The system described is a complex combination of several physical phenomena, leading to a highly nonlinear system and a large parameter space. Therefore, simulations using MATLAB's Simulink and Simscape are chosen to characterize the system. A simplified block diagram is illustrated (**Fig. 4**), and a detailed block diagram is provided in the Supplementary Material (**Fig. S1**). In essence, the nominal parameters in **Table 1** are input to the system and the outputs are tracked in real time. The sphere amplitude, $x_S$, is reported to gather a clear sense of the operation of the device, and the change in the coupled system resonant frequency, $\omega_{0C}$, is the ultimately proposed detection method. These values are gathered at some point in the time response after initial transients have settled (typically 0.35 s). We designed the simulation only with tools we typically use in the laboratory, to facilitate the translation to experimental measurement. A feedback approach is employed, where the sphere's frequency is detected by a zero crossing (negative slope). A waveform at double the frequency involving the Casimir coupling and a gradient magnetic field is fed back into the actuation of the sphere.

At the start of the simulation (t=0), we assume that the sphere is resonating at the unloaded resonance frequency, $\omega_{0S}$, and the magnet is resonating at $2\omega_{0S}$. The objects are proposed to incrementally approach each other from a large separation distance (> 1 µm, where the Casimir force is minimal) to reach the prescribed separation. This approach is based on the experimental observation[5] of the Casimir Force, using a similar platform. As the Casimir coupling begins to interact, our feedback system will adjust the actuation of the sphere (**Fig. 4**). The resonant mode of the system is pumped such that there is little amplitude decay due to damping, a technique[35] which is well-characterized in simulation and experimentally. Previous analytical work[12] and experimental work[30-31] have controlled the phase of objects in a parametrically amplified system, but this is challenging to do in dynamic simulation experiments, such as with Simulink.

This is ultimately why we choose a time delay approach, where the translation to experiment is straightforward with a precision digital delay generator (such as the the DG645, SRS). It is worth noting, however, that a constant time delay will result in a changing phase delay for a system with changing oscillation frequency (as is our case). Therefore, this design is notably different from a phase delay parametric pumping system.

A gradient magnetic field is introduced to the system in **Fig. 4** by a force, $F_M$, on the magnet. This results in a linear deflection of the magnet by Eq 6, changing $x_M$, and dynamically altering the separation, $x_{SM}$. This force is imposed statically in **Figs. 5** and **7**, and is imposed dynamically in **Fig. 6**. The sphere amplitude is reported to intuitively convey the behavior of the device, but the resonant frequency of the coupled system is measured to infer the measurand, a gradient magnetic field. As such, we define a sensitivity, $S_{freq}$ (Hz / pT/cm), shown in Eq 11, as the ratio between the natural frequency of the coupled system, $f_{0C}$, and the change in gradient magnetic field, $\nabla B_x$. In **Fig. 7**, this is computed by taking the slope between two consecutive data points.

$$S_{freq} = \frac{\Delta f_{0C}}{\Delta(\nabla B_x)} \quad (11)$$

The parametric pumping is expected to enhance the coupling of the Casimir Force. Analytically, the sphere-plate Casimir coupling was shown[12] to have a max gain proportional to the inverse of the 5th power of the separation when a phase delay of 0 degrees (or time delay of 0 s) is used and the sphere's amplitude is detected. To our knowledge, there is no experimental observation of a parametrically amplified Casimir coupling. Due to the anticipated danger of pull-in at maximum amplification, we identify a region of design space resilient to pull-in and with a parametrically amplified Casimir coupling. We leverage a shift in the coupled resonant frequency to propose a frequency detection scheme.

**Results: Magnetic Sensing with the MEMS Casimir Gradiometer**

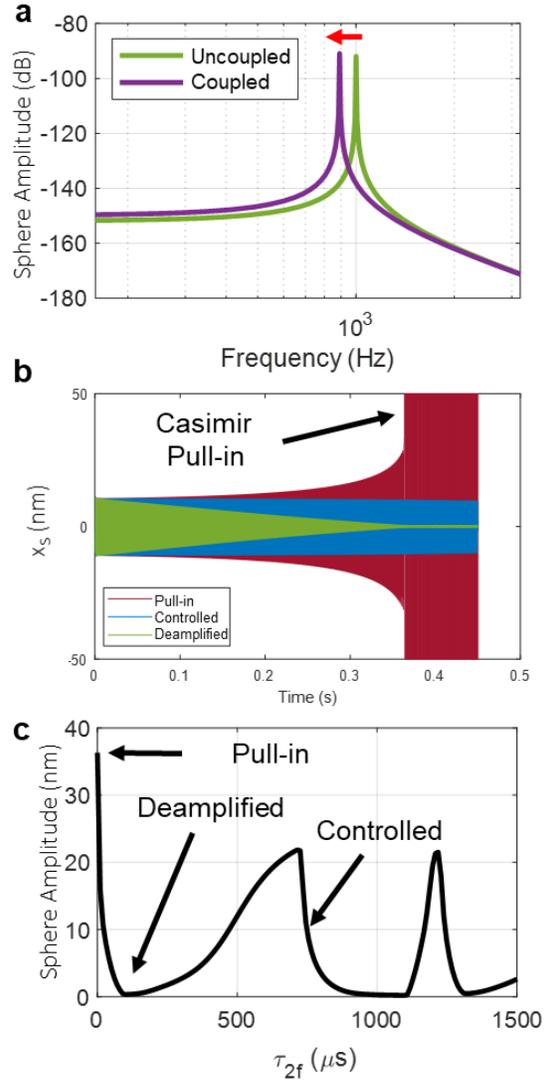

**Fig. 5: Characterized System Response.** (a) Bode plots of uncoupled ($x_{SM} > 1$ μm) and coupled configuration ($x_{SM}$ ~100nm). The quality factor is near 1000 in both cases and the resonant frequencies are 1000 and 850 Hz respectively. The decrease in frequency can be explained by the asymmetric interaction with the nearby plate, which imposes the attractive Casimir force. (b) and (c) are in the coupled configuration, with nominal parameters from **Table 1**, except $\tau_{2f}$. (b) A phase delay of the parametrically amplified magnet system can yield a highly unstable and sensitive response (pull-in), a stable and controlled response, and a deamplified response. (c) The sphere amplitude for a finely swept $\tau_{2f}$ (at 0.35s of the time response after transients have settled). The maximum is near a delay of zero and the profile is strongly nonlinear.

The system is characterized by simulation (**Fig. 5**), where results from the dynamics of the sphere are analyzed. The uncoupled and Casimir coupled systems are illustrated using Bode plots (**Fig. 5a**). As shown in **Table 1**, the resonant frequency of the unloaded system is designed to be 1 kHz with a Quality factor of 1000. The Casimir coupling causes a pronounced decrease in the resonant frequency (10s to 100s of Hz). The high quality factor of the system facilitates frequency shift detection. Recall that the uncoupled case is when the separation between the sphere and magnet is larger than 1 µm. At such separations, the sphere responds to a harmonic drive with symmetric oscillations shepherded by the spring's linear restoring force. The coupled case is when the separation between the magnet and sphere is in the range of 100 nm. In this regime, the spring's restoring force deviates from parabolic, becoming asymmetric (**Fig. 3**), leading to nonlinear dynamics.

In the coupled configuration, the parameters are tuned to characterize the system and investigate useful areas of design space. The first parameter of interest is the time delay of the magnet's oscillation, $\tau_{2f}$. In **Fig. 5b**, we show the temporal response to various time delays, labeled Pull-in ($\tau_{2f}=0$ µs), Controlled ($\tau_{2f}=750$ µs), and Deamplified ($\tau_{2f}=150$ µs). This characterizes three responses for different values of $\tau_{2f}$, and the responses to finely swept values is characterized (**Fig. 5c**) by the AC amplitude of the sphere after transients have settled (t=0.35 s). We find a maximum amplification at $\tau_{2f}=0$ µs (corresponding to a phase delay of 0 degrees), which is consistent with previous analytical analysis[12] of a phase delay system. While the $\tau_{2f}=0$ µs delay is the most sensitive region of the design space, we find that the two resonators in the Casimir coupled system always result in pull-in if left to interact for an arbitrarily long time. This indicates malfunction and destruction of the sensor. Consequently, we avoid this region of design space and investigate non-zero time delays for a sensitive, controlled condition. For non-zero time delays, however, this system is different than a phase delay system and may not be directly compared. We find a maximum deamplification at $\tau_{2f}=150$ µs followed by two other maxima (**Fig. 5c**). After the peak at $\tau_{2f}=1200$ µs, the pattern repeats with slightly decreased amplitudes due to energy lost per cycle (not shown). At $\tau_{2f}=750$ µs, we report a stable, controlled oscillation of the sphere (**Fig. 5b**), which is expanded on in **Fig. 6** and **7**.

The time response of the controlled system ($\tau_{2f}=750$ µs) is dynamically characterized over an elapsed time of 2 s for zero gradient magnetic field input (**Fig. 6a**) and for a slowly varying sine wave oscillation of a gradient magnetic field (**Fig. 6b**). The resonant frequency of the coupled system in these two conditions, $f_{0C}$, is tracked in both configurations (**Fig. 6c**). The coupled system with no input field reaches a steady state after approximately 200 ms, after which point the signal is stable. A dynamic gradient field input of a 1 Hz sine wave, 4 pT/cm peak-to-peak is chosen to demonstrate the response of the system to a slowly varying magnetic field. For a gradiometer[27] sensitivity of 1 µV/(fT/cm), and a magnet spring constant of 25 mN/m (40 times softer than the experimental platform in ref. 27), this gradient field would yield a 1 nm peak-to-peak oscillation of the magnet. The system responds with a 1 Hz oscillation of approximately 4 Hz peak-to-peak shift in $f_{0C}$. For very small gradients, the response will be nearly linear, and for larger gradient fields, the response will be a nonlinear, asymmetric sine wave. However, the change in the coupled resonant frequency directly maps to a change in the gradient, and so the true gradient signal can be easily calculated by ratiometric conversion.

The sensitivity, $S_{freq}$, is tunable with respect to separation (**Fig. 7**). System parameters of $\tau_{2f}=750$ µs and $A_{2f}=1$ nm are selected. For changing separation $x_{SM}$, the change in $f_{0C}$ is recorded on the first y-axis (right). Sensitivity, on the second y-axis (left), is calculated using Eq 11. The gradient field used to calculate sensitivity is the equivalent field that would deflect the magnet, altering $x_{SM}$. An inverse power function profile is observed as a result. Such a profile is expected, as the Casimir Force follows an inverse cubic function with respect to separation (Eq 7). This design is not static, however, and the parametric pumping (Eq 8) modifies the dynamics. As introduced earlier, analytical work[12] with an equivalent time delay of 0 s proposed detection proportional to the 5$^{th}$ power of the separation. We find the region with zero delay to lead to pull-in (**Fig. 5b**), and so utilize the controlled response found at $\tau_{2f}$=750 µs. To compare to the previously proposed detection scheme at maximum amplification, we use an inverse exponential fit, $y = a/(x-b)^c + d$, where $a$, $b$, $c$, and $d$ are constants and $c$ describes the power relationship. Fitting the data in **Fig. 7** yields c=2.6, for our proposed frequency detection scheme. While this is significantly less sensitive than the 5$^{th}$ power relationship (c=5), the Casimir coupling is amplified and resilient to pull-in.

The best-case resolution of the system is discussed for frequency shift detection of the high quality peak. For laboratory based frequency detection systems, such as Agilent's 53132A frequency counter, a resolution of 10 parts-per-million using a 1 s gate time would be relatively standard. A maximum sensitivity, $S_{freq}$, of 6 Hz/(pT/cm) is observed (**Fig. 7**). Therefore, a frequency detection scheme would have a best-case resolution of 1.6 aT/cm at 1 Hz. While we anticipate thermo-mechanical noise to experimentally limit the technique at a much larger gradient magnetic fields, this would constitute 4 orders of magnitude improved resolution (10,000 fold) for the best-case scenario of the presently designed MEMS single-point gradiometer[27].

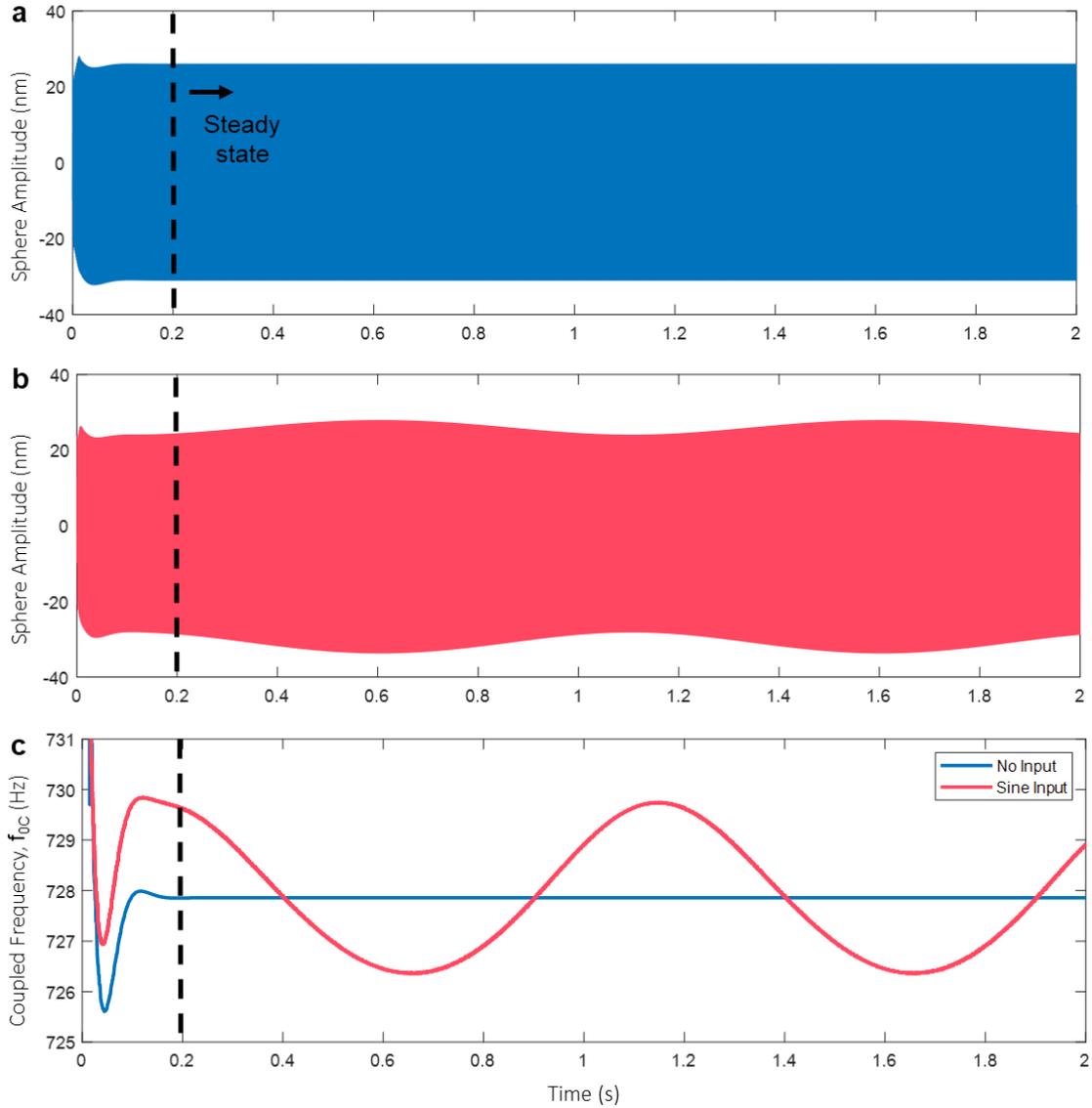

**Fig. 6: Response to Slowly Varying Gradient Magnetic Field.** The controlled configuration at $\tau_{2f}$=750 μs is tracked for 2 s following an imposed initial condition. (a) shows the sphere amplitude response to no magnetic field input, where the two resonators are set to interact and reach a steady state after about 200 ms. (b) shows the sphere amplitude response when the magnet is modulated by a 1 Hz, 4 pT/cm peak-to-peak gradient magnetic field. (c) tracks the shift in the coupled resonance frequency, $f_{0C}$, in both conditions (a) and (b). A 1 Hz oscillation is observed in both (b) and (c), where tracking the frequency shift of the high quality peak reports a clean signal. For small deflections of the magnet (1 nm peak-to-peak here), the response is nearly linear, but larger deflections will result in an asymmetric sine wave.

While this work focuses on employing a Casimir-driven parametric amplifier to MEMS sensing, there are other configurations which may be useful to consider. The electrostatic force acts at larger separations (> 100 nm), which may be more resilient to pull-in. Electrostatic forces may also be parametrically amplified[30-31], and are also nonlinear (albeit less sensitive than the Casimir Force). Furthermore, it has been discussed that sensitivity varies with separation, and therefore with gradient magnetic field input. Although complicated for a highly dynamic system, we suggest the investigation of a null-sensing technique which may enable the device to sit at a single sensitivity. Using an additional feedback mechanism may be able to control the center-positions of both oscillators (keeping them constant) and may afford control of a constant, high sensitivity, such as the 6 Hz/(pT/cm) reported here.

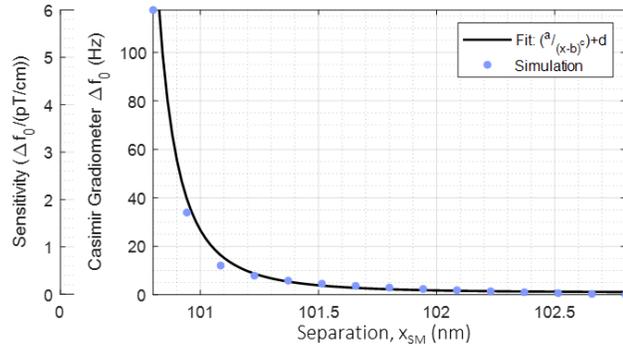

**Fig. 7: Tunable Sensitivity to Separation.** The separation, $x_{SM}$, is varied in a 3 nm range, resulting in an inverse exponential profile in frequency shift of the coupled resonance, $f_{0C}$. The sensitivity, $S_{freq}$, is calculated using a gradient magnetic field that would cause an equivalent deflection in $x_{SM}$. The maximum sensitivity reported is 6 Hz/(pT/cm). Standard frequency detection with 10 ppm and 1 s gate time would yield a best-case resolution of 1.6 aT/cm. While we anticipate limitations due to thermo-mechanical noise, this over a 10,000 fold improvement on the previous design[27] of the MEMS single-point gradiometer. While a zero delay configuration may provide maximum amplification, we report a highly sensitive design using a $\tau_{2f}$ = 750 µs delay, which is resilient to Casimir pull-in.

Resonant MEMS devices such as this design are often limited by several types of noise. The characteristic 1/f noise from mechanical and electrical sources will largely not affect the resonator coupling near 1kHz, but the low frequency changes in separation which are sought to be measured here will likely be affected. We suggest that techniques such as chopper stabilization and lock-in amplification be employed to reduce this effect. Low frequency magnetic noise, such as from power lines in an urban area or the Earth's magnetic field will interfere with sensitive magnetic measurements. As was experimentally analyzed previously[27], this interference would present itself as a torque on the magnet, equal to the cross product of the magnetic moment and interference field (in plane or out-of-plane). Standard gradiometer designs[23-24] reduce this noise from distant sources by subtracting the signals from two closely spaced magnetic sensors. This spacing is typically on the order of 1 cm, and so, by design, our system will improve this reduction with a spatial element 0.25 mm in length along the sensitive axis. Furthermore, our technique presents a subtractionless measurement, offering a reduction to associated error for gradiometric measurements. Although a central goal of our technique is to develop a sensor for unshielded biomagnetic measurements, shielding may still be employed to further reduce the interference of low frequency magnetic fields.

Our design is intended for an ambient temperature and pressure environment, and so we anticipate thermomechanical noise to be a dominant influence on our measurements. Thermal damping on each of the resonators is mitigated in part by the sine-wave feedback pumping. Our simulations did not investigate noise directly, but the design and experimental implementation may be directly compared to previous work[27]. Squeeze film damping has been shown[36] to be common on MEMS devices with gaps smaller than

5 μm, such as those designed in this work. Although we intend for this design to be used in ambient environments, vacuum packaging or cryogenic environments would further reduce the effects of damping. Finally, it was theoretically shown[37] that another source of damping for a dynamic Casimir oscillator may arise from the nonuniform relative acceleration of the sphere and plate, which enclose the nonlinear properties of vacuum. It may be interesting to combine our sensitive platform design with cryogenic and magnetically shielding environment to investigate this effect experimentally. We are confident that the design strategies presented in this work, in addition to those that reduce the effects of noise will profoundly enhance the performance of single-point MEMS gradiometers.

**Conclusion**

We have investigated a quantum-derived coupling of two resonant microstructures to achieve extremely high sensitivity to changes in a gradient magnetic field. The resonators are coupled by the nonlinear Casimir Force, which arises from the electromagnetic interaction between closely spaced dielectrics (near 100nm) in a sphere-plate geometry. A customized parametric amplification technique is developed, where one resonator is synchronized at double the frequency of the other, and the time delay is tuned to find a steady state solution. The frequency shift of the high quality, coupled resonance peak is detected to infer a measured gradient magnetic field. A slowly varying field at 1 Hz is imposed, where a best-case resolution is calculated to be 1.6 aT/cm at a sensitivity of 6 Hz/pT/cm. This is a 10,000 fold improvement on the best-case resolution of the previously designed MEMS single-point gradiometer. Many applications, especially the measurement of biomagnetic fields, already rely on complex quantum metrology. The MEMS quantum-enhanced gradiometer presented in this work paves a path toward unshielded, ambient temperature measurements of extremely weak gradient magnetic fields.

**Acknowledgements**

We would like to thank Alex Stange for helpful consultations regarding state-of-the-art Casimir metrology and Nicholas Fuhr for productive discussions pertaining to magnetic sensing and other metrology. This work was supported by the NSF CELL-MET ERC award no. 1647837 and a SONY Faculty Innovation Award.

**Conflict of Interests**

The authors declare that they have no conflict of interest.

**Author Contributions**

The design was conceived and refined by D.J.B., D.K.C., M.I., J.J. and Z.Y. The simulation was done by J.J. and Z.Y. with assistance from D.J.B. All data were collected, interpreted, and analyzed by J.J. and Z.Y. The manuscript was written by J.J. with input from Z.Y. and edited by all authors.